\documentclass[numbers=noenddot]{jfm}
\usepackage[latin9]{inputenc}

\usepackage{array, ragged2e}
\usepackage{float}
\usepackage{mathtools}
\usepackage{graphicx}
\usepackage{esint}
\usepackage{verbatim}
\usepackage{caption}
\usepackage{subcaption}
\renewcommand\[{\begin{equation}}
\renewcommand\]{\end{equation}}

\numberwithin{equation}{section}
\usepackage{amsmath} 
\newcommand{\beq}{\begin{equation}}
\newcommand{\eeq}{\end{equation}}
\newcommand{\lb}{\left(}
\newcommand{\rb}{\right)}
\usepackage{tikz}
\usepackage{pgfplots}
\usepackage{overpic}
\usepackage{hyperref}
\usepackage{todonotes}
\author{Graham P. Benham\aff{1}\corresp{\email{benham@maths.ox.ac.uk}}}
\affiliation{\aff{1} Mathematical Institute, University of Oxford, Oxford OX2 6GG, United Kingdom}

\begin{document}

\title{The near-field shape and stability of a porous plume}

\maketitle

\abstract{
When a fluid is injected into a porous medium saturated with an ambient fluid of a greater density, the injected fluid forms a plume that rises upwards due to buoyancy. In the near-field of the injection point the plume adjusts its speed to match the buoyancy velocity of the porous medium, either thinning or thickening to conserve mass. These adjustments are the dominant controls on the near-field plume shape, rather than mixing with the ambient fluid, which occurs over larger vertical distances. In this study, we focus on the plume behaviour in the near-field, demonstrating that for moderate injection rates the plume will reach a steady state, whereby it matches the buoyancy velocity over a few plume width-scales from the injection point. However, for very small injection rates an instability occurs in which the steady plume breaks apart due to the insurmountable density contrast with the surrounding fluid. The steady shape of the plume in the near-field only depends on a single dimensionless parameter, which is the ratio between the inlet velocity and the buoyancy velocity. A linear stability analysis is performed, indicating that for small velocity ratios an infinitesimal perturbation can be constructed that becomes unstable, whilst for moderate velocity ratios the shape is shown to be stable. Finally, we comment on the application of such flows to the context of CO$_2$ sequestration in porous geological reservoirs. 
 }

\section{Introduction}


Buoyant plumes in porous media may result from thermally-driven convection or during injection scenarios involving fluids of different densities. Such flows are relevant within the context of numerous environmental and geophysical applications, such as groundwater contaminant transport due to waste leakage \citep{macfarlane1983migration}, geothermal power production \citep{woods1999liquid} and the geological storage of CO$_2$ emissions in subsurface reservoirs \citep{huppert2014fluid}. 
Whilst such plumes have been studied in some detail far away from their origin, few studies have investigated the near-source behaviour. 
In particular, it is not known how the shape of a porous plume evolves close to the point of its formation, nor whether this shape remains in a stable state or if it breaks apart due to instabilities.
However, it is important to understand the characteristics of the near-field plume due to the effects it can have on the pressure near the injection point and consequent flow rates \citep{gilmore2022leakage}.

A wide body of literature has been developed surveying different buoyant flows in porous media.  These include studies of convective instabilities in a Rayleigh-B\'{e}nard cell \citep{graham1994plume}, convective shutdown behaviour \citep{hewitt2013convective}, the onset and evolution of convective fingers \citep{wooding1997convection1,wooding1997convection2}, and mixing effects during injection into a porous medium \citep{lyu2016experimental}.
It has been demonstrated that a quasi-steady regime exists in both two and three dimensional Rayleigh-B\'{e}nard cells in which convection occurs in columnar structures \citep{hewitt2013stability,hewitt2017stability}. 
For buoyant flows which are not thermally driven, but are instead driven by injection, similar columnar structures have been observed. For example, \citet{gilmore2022leakage} described the behaviour of a two-dimensional buoyant column of fluid with weakly varying thickness, resulting from leakage through an impermeable baffle. In that study it was shown that the shape of the column affects the near-baffle pressure and consequent leakage rates, indicating the need to model such scenarios accurately. However, there is no study which describes the generic shape of a porous plume near its source, nor the criteria for which this remains stable.

This study describes such a porous plume supplied by a constant injection (i.e. not thermally-driven), focusing on its shape and stability in the near-field (within a few plume width-scales) of its origin. We ignore the effects of mixing with the ambient fluid, since these occur over much greater length scales and are described by other studies \citep{sahu2015filling,lyu2016experimental}. We establish the criteria for the existence of a steady state regime and we demarcate the parameter values for which this becomes unstable. 
In particular, if the injected fluid is supplied with a velocity much smaller than the buoyancy velocity (i.e. the equilibrium rise speed within the porous medium), the interface separating the plume from the ambient fluid becomes unstable at a critical distance downstream. On the other hand, if the inlet velocity is sufficiently close to the buoyancy velocity, this  instability is suppressed and the plume maintains a steady shape.

The structure of the paper is laid out as follows. In Section 2 the flow scenario is described for both two-dimensional plumes resulting from a line source, and axisymmetric plumes resulting from a circular source, deriving both analytical and numerical solutions in the steady state. Comparisons are also made with the porous media tank experiments of \citet{gilmore2022leakage}. Section 3 treats the stability of these steady plume shapes using a linear perturbation analysis. Finally, Section 4 closes with some concluding remarks and discusses the possible application of our results to injection scenarios during CO$_{2}$ sequestration.

\section{Porous plumes in the near-field of injection}
\label{sec_plumeintro}

\begin{figure}
\centering
\begin{tikzpicture}[scale=0.6]
\node at (0,-5.) {\includegraphics[width=0.81\textwidth]{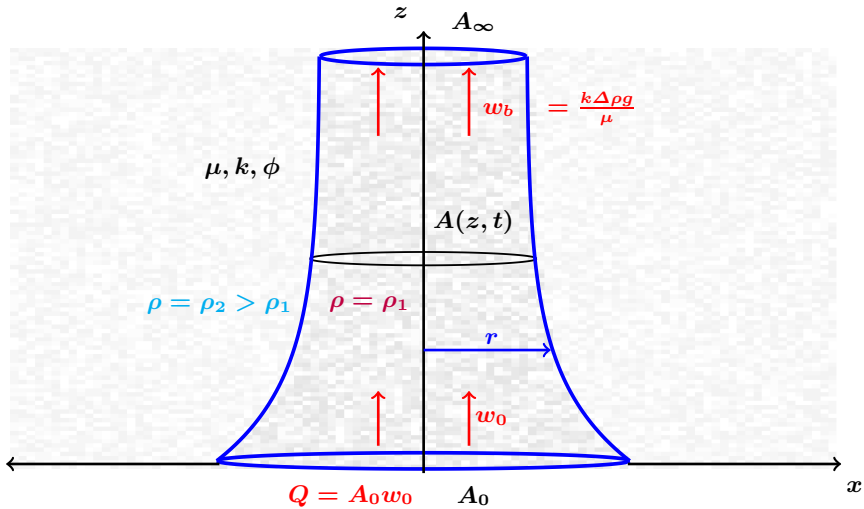}};
\draw[line width=1,black,->] (4.5,-9.5) -- (9.2,-9.5);
\draw[line width=1,black,->] (-4.5,-9.5) -- (-9.2,-9.5);
\draw[line width=1,black,->] (0,-9.7) -- (0,0);
\draw[line width=1,red,<-] (1,-0.8) -- (1,-2.3);
\draw[line width=1,red,<-] (-1,-0.8) -- (-1,-2.3);
\node[red] at (3.0,-1.7) { $\boldsymbol{w_b\quad=\frac{k\Delta \rho g}{\mu}}$};
\draw[line width=1,red,<-] (1,-7.9) -- (1,-9.1);
\draw[line width=1,red,<-] (-1,-7.9) -- (-1,-9.1);
\node[red] at (1.5,-8.5) { $\boldsymbol{w_0}$};
\node[black] at (1.1,0.2) { $\boldsymbol{A_\infty}$};
\node[black] at (1.1,-10.2) { $\boldsymbol{A_0}$};
\node[red] at (-1.6,-10.2) { $\boldsymbol{Q=A_0 w_0}$};
\node[purple] at (-1.2,-6) { $\boldsymbol{\rho=\rho_1}$};
\node[cyan] at (-4.5,-6) { $\boldsymbol{\rho=\rho_2>\rho_1}$};
\node[black] at (-4,-3) { $\boldsymbol{\mu,k,\phi}$};
\node at (9.5,-10) { $\boldsymbol{x}$};
\node at (-0.5,0.4) { $\boldsymbol{z}$};
\node[black] at (1.1,-4.2) { $\boldsymbol{A(z,t)}$};
\draw[line width=1,blue,->] (0,-7) -- (2.8,-7);
\node[blue] at (1.5,-6.75) { $\boldsymbol{r}$};
\end{tikzpicture}
\caption{Schematic diagram of the flow scenario in the case of a circular source. The injected fluid $z\geq0$ is fed by a flow $w_0$ through a disk region of area $A_0$, and speeds up to match the natural buoyancy velocity $w_b$ downstream. Hence, the plume cross-section $A(z,t)$ thins out from $A_0$ to $A_\infty$ to conserve  mass. \label{schem}}
\end{figure}

We consider the constant injection $Q$ of a fluid of density $\rho_1$ into an infinite porous medium saturated with a heavier fluid of density $\rho_2>\rho_1$, as illustrated in figure \ref{schem}\footnote{Note, this study also applies to the configuration of a heavier fluid injected into a lighter fluid $\rho_2<\rho_1$, due to the Boussinesq approximation \citep{soltanian2016critical,amooie2018solutal}, in which case figure \ref{schem} is inverted.}. 
For simplicity, we assume that the fluids have the same viscosity $\mu_1=\mu_2=\mu$.
Since the injected fluid is lighter than the ambient fluid it rises upwards, forming an ascending plume of cross-sectional area $A(z,t)$.

There are two spatial regimes characterised by $A_0=A(0,t)$, the area at the point of injection. 
In the near-field regime $z=\mathcal{O}(A_0^{1/2})$, which is the focus of the current study, the plume adjusts its shape to conserve mass whilst matching the equilibrium buoyancy velocity of the porous medium (i.e. buoyancy balancing viscous resistance) and the effects of mixing with the ambient fluid are negligible. 
Over much greater length scales $z\gg A_0^{1/2}$ the injected fluid mixes with the ambient fluid causing the buoyancy to decrease and the width of the plume to increase as the flow moves upwards.
For example, the experiments of \citet{sahu2015filling} revealed that plume width changes due dispersive mixing occur over vertical length scales of $z\sim \mathcal{O}(10 A_0^{1/2})$.

Therefore, in the current study we neglect the effects of mixing and focus only on the changes in the plume shape due to mass conservation as it adjusts its velocity. Hence, we treat the injected and ambient fluids as immiscible, such that the interface between them remains sharp (e.g. see sharp interface models of other gravity-driven flows \citep{huppert1995gravity}). 
We consider both the case of injection from a line source, in which the resultant flow only varies in the horizontal ($x$) and vertical ($z$) directions, and injection from a circular source, in which the resultant flow is axisymmetric and varies with cylindrical coordinates ($r$ and $z$), as shown in figure \ref{schem}.

 If the injection flow rate $Q$ is sufficiently small, we expect an instability to occur in which the shape of the plume $A(z,t)$ becomes unsteady due to the density contrast of a heavier fluid sitting above a lighter fluid \citep{lord1900investigation,taylor1950instability}. Likewise, we expect another regime corresponding to larger flow rates in which the shape remains steady  $A=A(z)$. The aim of this study is to first describe the steady state regime for the near-field plume, and then to address the criteria for the stability of this steady state.

\subsection{Steady plumes}

To address the steady state regime, we describe the shape of the plume $A(z)$ above the injection height ($z\geq0$) in general terms that apply to both linear and circular sources. 
At the injection point the vertical inflow velocity is 
\beq
w_0=\frac{Q}{A_0}.\label{inflow}
\eeq
Likewise, the buoyancy velocity of the injected fluid, which is the equilibrium rise speed in the porous medium (i.e. buoyancy balancing viscous resistance), is given by
\beq
w_b=\frac{k\Delta\rho g}{\mu},\label{buoy}
\eeq
where $k$ is the permeability of the medium and $\Delta \rho=\rho_2-\rho_1$. Due to mass conservation the far-field\footnote{Note, we use the term \textit{far-field} here and throughout the manuscript to refer to the length scale over which the plume velocity approximately matches the buoyancy velocity. This is not to be confused with the even greater length scales over which the effects of mixing are important (since these are not studied here).} cross-section is given by $A_\infty=Q/w_b$.
Hence, the key dimensionless parameter in this study is the ratio between the inlet and buoyancy velocities,
\beq
W=\frac{w_0}{w_b}.
\eeq
As the flow moves downstream (i.e. upwards) the plume must become thinner ($A_\infty<A_0$) for sub-buoyancy velocities $W<1$ and thicker ($A_\infty>A_0$) for super-buoyancy velocities $W>1$.

The flow within the injected fluid is governed by the Darcy equations
\begin{align}
\nabla\cdot \mathbf{u}&=0,\label{darcy1}\\
\mathbf{u}&=-\frac{k}{\mu}\lb \nabla p + \rho_1 g \hat{\mathbf{k}}\rb,\label{darcy2}
\end{align}
where $\mathbf{u}$ is the Darcy velocity vector and $p$ is the pressure. 
The flow in the ambient fluid is only coupled to the injected fluid via the boundary conditions, which we discuss in the next section. Hence, for the purposes of this study we omit further details of how to model the ambient flow outside the injected region, since the behaviour of the injected flow is of primary interest.

The Darcy equations \eqref{darcy1}-\eqref{darcy2} are accompanied by boundary conditions that take a different form depending on whether the flow is injected from a line source or a circular source. Hence, we address the former and latter cases separately in Sections \ref{subsec_line} and \ref{subsec_rad}.

\subsection{Plume shape: The case of a line source}
\label{subsec_line}

 In the case of the line source (with $A(z)=a(z)d$, where $d$ is the depth in the third dimension) we impose boundary conditions within the injected fluid region of the form
\begin{align}
u&=0:&x=0,\label{bcfirst}\\
w&=w_0:&z=0,\\
w&\rightarrow w_b:&z\rightarrow \infty,\\
u&=wa'(z):& x=a(z),\label{bcpen}\\
p&=p_a-\rho_2 g z:&x=a(z),\label{bclast}
\end{align}
where $p_a$ is the ambient hydrostatic pressure at $z=0$. The above boundary conditions correspond with imposing symmetry on the $z$ axis, constant inflow at the source, matching with the far-field buoyancy velocity, and applying the kinematic and dynamic conditions at the sharp interface, respectively. The above system \eqref{darcy1}-\eqref{bclast} is a free boundary problem for both the flow and the shape of the interface $a(z)$. To proceed, we seek a solution of the form
\begin{align}
p&=p_a-\rho_2 g z +\hat{p},\label{form1}\\
u&=\hat{u},\\
w&=w_b+\hat{w},\label{form3}
\end{align}
such that the hatted variables satisfy the new system of equations
\begin{align}
\nabla^2\hat{p}&=0,&\label{redfirst}\\
\hat{p}_x&=0:&x=0,\\
\hat{p}_z&=\Delta\rho g(1-W):&z=0,\\
\hat{p}_z&\rightarrow 0 :&z\rightarrow \infty,\label{redpen}\\
\hat{p}_x&=(\hat{p}_z-\Delta \rho g)a'(z):&x=a(z),\\
 \hat{p}&=0:&x=a(z),\label{redlast}
\end{align}
where subscripts denote partial derivatives. In the case where the inlet velocity is close to the buoyancy velocity ($W\approx 1$) the solution to \eqref{redfirst}-\eqref{redlast} was calculated by  \citet{gilmore2022leakage} using the method of separation of variables, in which case
\beq
\hat{p}\approx -(1-W)\frac{8\Delta \rho g a_0}{\pi^2}\sum_{n=0}^\infty \frac{(-1)^n}{\lb 2n+1\rb^2} \cos\left[\lb 2n+1\rb{\pi x}/{2a_0}\right]e^{-\lb 2n+1\rb{\pi z}/{2a_0}},\label{gilmoreapprox}
\eeq
where $a_0=a(0)$. Likewise, the approximate plume shape is given by the solution to
\beq
a'(z)\approx -(1-W)\frac{4}{\pi}\tanh^{-1}\left[ e^{-\pi z/2 a_0} \right].\label{plumedif}
\eeq

\begin{figure}
\centering
\begin{tikzpicture}
\node at (0,0) {\includegraphics[height=0.3\textheight]{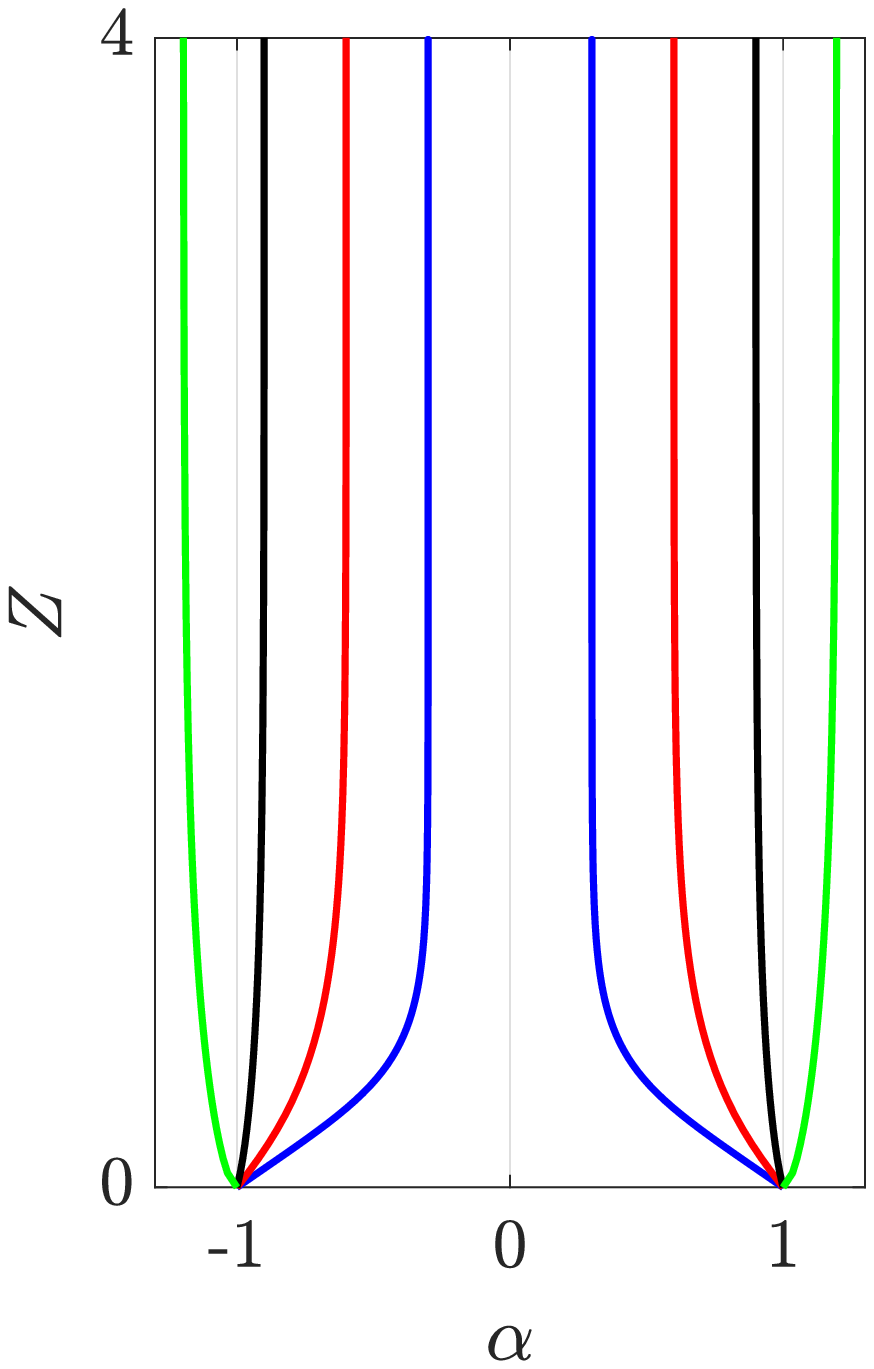}};
\node at (7,0) {\includegraphics[height=0.3\textheight]{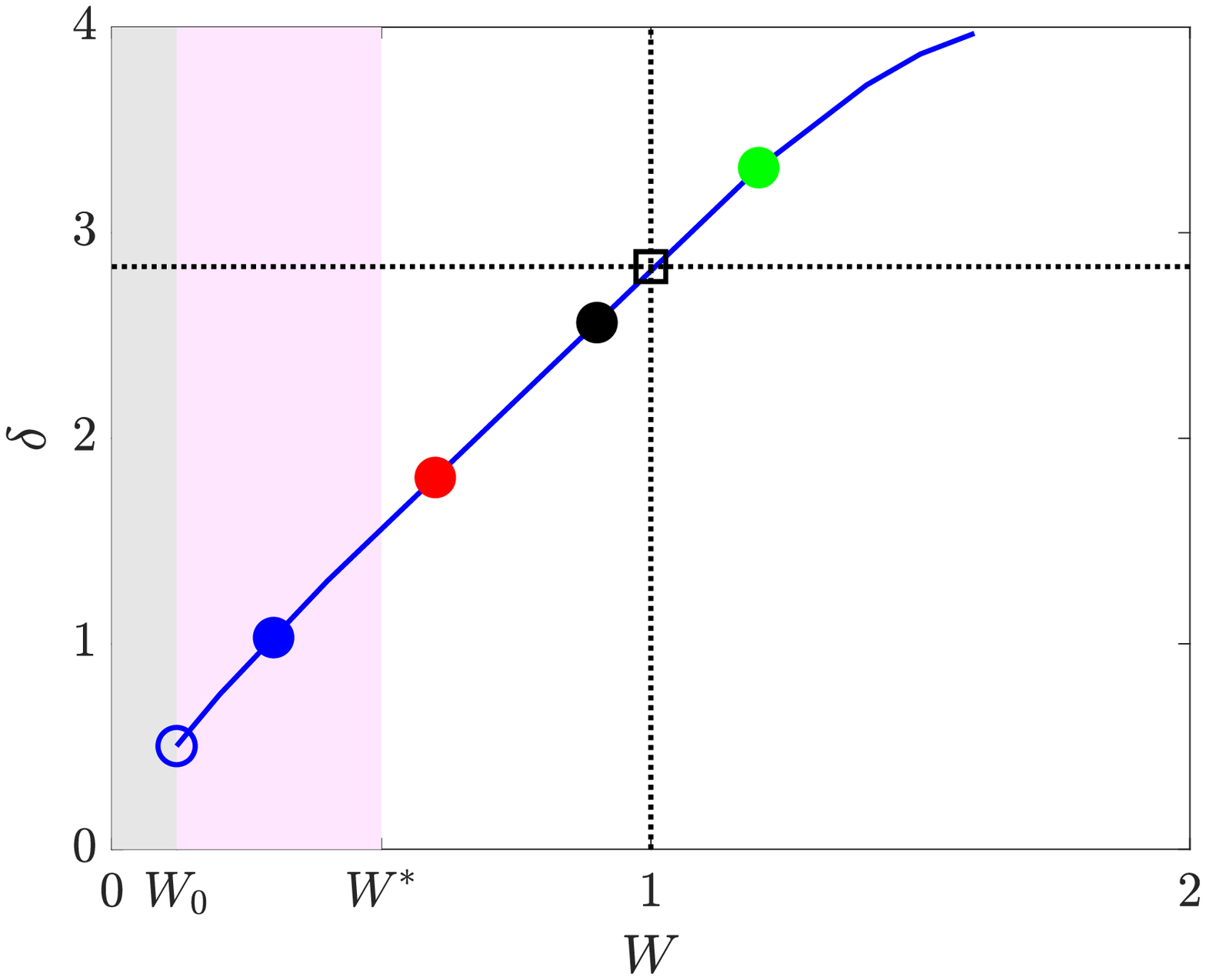}};
\node at (9,0.8) {\scriptsize Stable};
\node at (4.9,0.8) {\scriptsize Unstable};
\node at (0.25,3.5) {(a)};
\node at (7,3.5) {(b)};
\node[blue] at (0.5,-1.6) {$\delta$};
\draw[blue, line width =1,<->] (0.35,-2.3) -- (0.35,-0.9);
\end{tikzpicture}
\caption{Numerical and analytical results for a thinning/thickening plume resulting from a line source. (a) Plume shape for different velocity ratios $W=w_0/w_b$ and (b) $99\%$ boundary layer distance $\delta$ as defined in \eqref{deltadef}. Critical values $W_0$ and $W^*$ are related to the sign of the determinant $\Delta$ \eqref{detfun} and the stability/existence of a steady solution. Dotted lines indicate the asymptotic limit $W\rightarrow 1$, for which the plume shape is given by the solution to \eqref{plumedif}. \label{shape}}
\end{figure}

In the case where the inlet velocity and the buoyancy velocity are not similar ($W\not\approx 1$), a numerical method must be used to calculate the solution. By converting to a set of scaled dimensionless variables
\beq
X=x/a(z),\quad Z=z/a_0,\quad P(X,Z)=\hat{p}(x,z)/\Delta \rho g a_0,\quad \alpha(Z)= a(z)/a_0,\label{stretch_coord}
\eeq
Laplace's equation \eqref{redfirst} becomes
\beq
\left[\alpha^{-2}\partial_{XX}+\lb\partial_Z-X\alpha'\alpha^{-1} \partial_X\rb^2\right]P=0,\label{lapdim}
\eeq
whilst the remaining boundary conditions become
\begin{align}
P_X&=0:&X=0,\label{dimfirst}\\
P_Z-X\alpha'\alpha^{-1} P_X&=1-W:&Z=0,\\
P_Z-X\alpha'\alpha^{-1} P_X&\rightarrow 0:&Z\rightarrow \infty,\label{dimmid}\\
P_X&=-\alpha\alpha'- {\alpha'}^2P_X :&X=1,\label{kindim}\\
 P&=0:&X=1.\label{dimlast}
\end{align}
The nonlinear system \eqref{lapdim}-\eqref{dimlast} is solved using Newton's method in combination with a finite difference scheme. The domain is discretised using a rectangular grid, with $X\in[0,1]$ and $Z\in[0,H]$, where the boundary condition \eqref{dimmid} is approximated at a large but finite value of $H=10$. We calculate the solution for a variety of values of $W$ (the only dimensionless parameter of the problem), using an $8^\mathrm{th}$ order finite difference scheme with a grid of $20\times 200$ points in the $X,Z$ directions. By employing the method of continuation using incremental changes in $W$, Newton's method converges in approximately $3$ steps (for each increment).

We plot examples of the shape $\alpha(Z)$ in figure \ref{shape}a for $W=0.3,0.6,0.9,1.2$. Thinning plumes are observed for $W<1$ whereas thickening plumes are observed for $W>1$, as expected.
One salient feature of the analysis is the distance over which the plume approaches its far-field width $a_\infty$. We define the $99\%$ {\it boundary layer} distance $\delta$ (dimensionless) as 
\beq
\left|\frac{\alpha(\delta)-W}{1-W}\right|=0.01,\label{deltadef}
\eeq
and this is plotted in figure \ref{shape}b for different values of $W$. We find that $\delta$ is  monotone increasing in the range $W\in [0.12,1.60]$.
For $W<0.12$, our numerical method fails to converge to a real-valued solution (which we discuss shortly), so no data is plotted. 
We also compare these values of $\delta$ with the analytical value in the case where $W$ is close to $1$ (e.g. via solution to \eqref{plumedif}).
In this case, the boundary layer distance is independent of $W$ at leading order, and is given by the approximate value $\delta\approx 2.83$ (see dotted lines in figure \ref{shape}b). 

Due to the kinematic boundary condition \eqref{kindim}, for the solution to remain real valued we require a non-negative determinant 
\beq
\Delta(Z):=\alpha^2-4\left.P_X\right|_{X=1}^2\geq 0,\label{detfun}
\eeq
for all values of $Z$. By writing the pressure gradient as a dimensionless velocity $P_X=-U$, we see that \eqref{detfun} can be interpreted as a balance criterion between the width of the plume $\alpha$ and the horizontal velocity required to sustain that width, $U$. For example, a thinning plume ($U<0$) with a shape that tapers smaller than a thickness $\alpha<-2U$ is not permitted by \eqref{detfun}. In general, whenever \eqref{detfun} cannot be satisfied this indicates that a steady plume shape is not possible.

The behaviour of the determinant function \eqref{detfun} depends on the value of the velocity ratio $W$. There are three solution regimes defined by two values of the velocity ratio, $W^*\approx 0.5$ and $W_0\approx 0.12$.
For velocity ratios $W>W^*$, the determinant is strictly positive $\Delta>0$ for all values of $Z$. For velocity ratios in the range $W_0<W<W^*$ the determinant is non-negative $\Delta\geq0$, but equals zero at some critical distance $Z=Z^*$ downstream of the inflow. 
For $0<W<W_0$, Newton's method fails to converge to a real-valued solution, indicating that a steady solution may not exist. 
The three different solution regimes are illustrated with shading in figure \ref{shape}b.
It should be noted that the sign of the determinant is closely linked with the stability criteria for the plume, and we discuss this later in Section \ref{instabsec}.


\subsection{Comparison with experiments}

In this section we compare our results for the steady plume shape (in the case of a line source) to the porous bead experiments of \citet{gilmore2022leakage}. These experiments were conducted in a thin rectangular tank of dimensions $40\times70\,\mathrm{cm}$ in the $x,z$ directions and $1\,\mathrm{cm}$ thick in the transverse ($y$) direction. The tank was filled with $3\,\mathrm{mm}$ ballotini beads and initially saturated with fresh water. Salty water dyed with red food colouring was injected into the top of the tank, using different salt concentrations to modulate the density contrast. 
Since salty water is heavier than fresh water, their experiments resulted in a falling plume rather than the rising plume studied at present. Therefore, we have inverted their experimental photos for comparison with our model. 
The inverted system behaves in approximately the same way as the current system due to the Boussinesq approximation \citep{soltanian2016critical,amooie2018solutal}.

\begin{figure}
\centering
\begin{tikzpicture}[scale=1.5]
\node at (0.5,0) {\includegraphics[height=0.25\textheight]{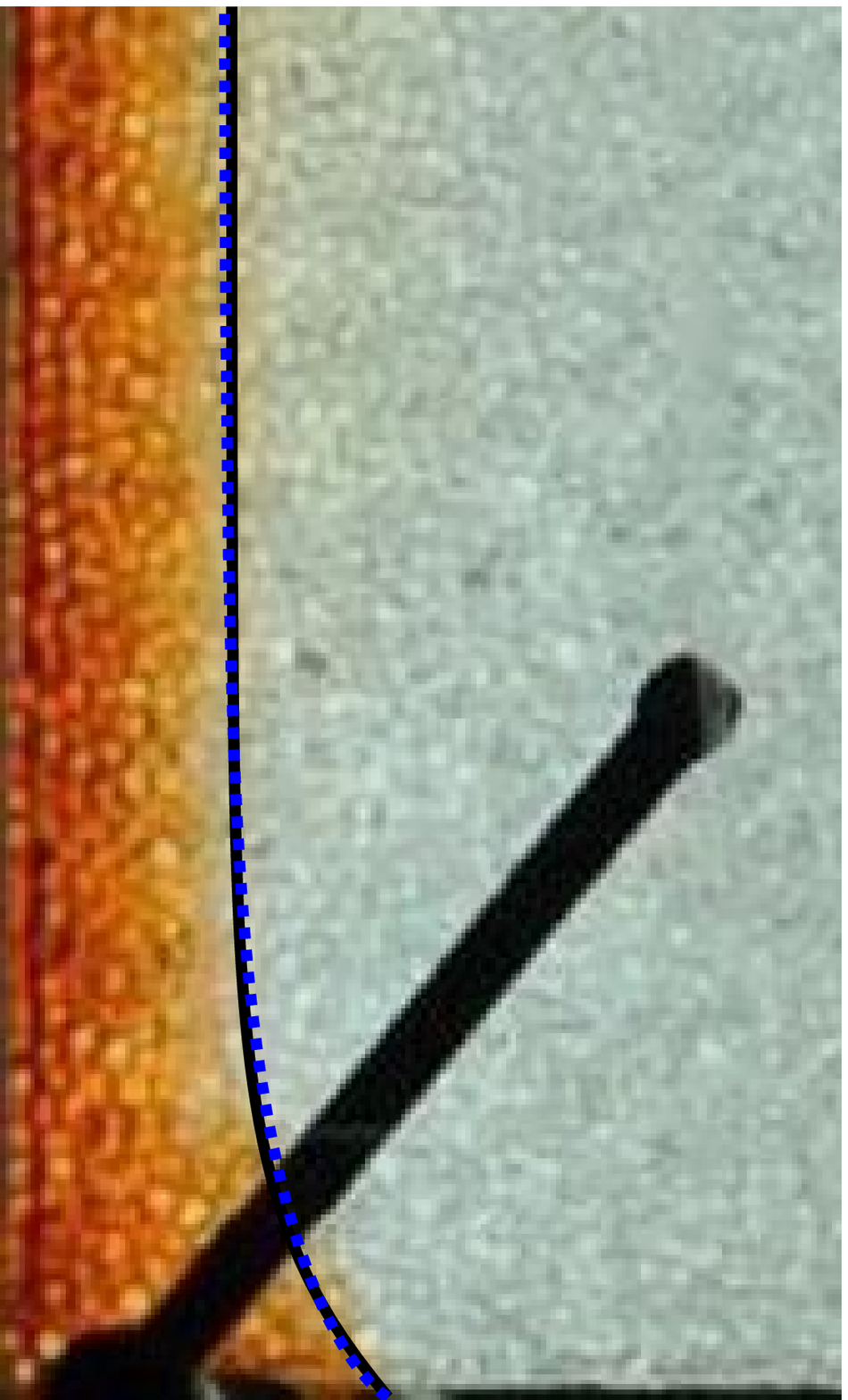}};
\node at (4,0) {\includegraphics[height=0.25\textheight]{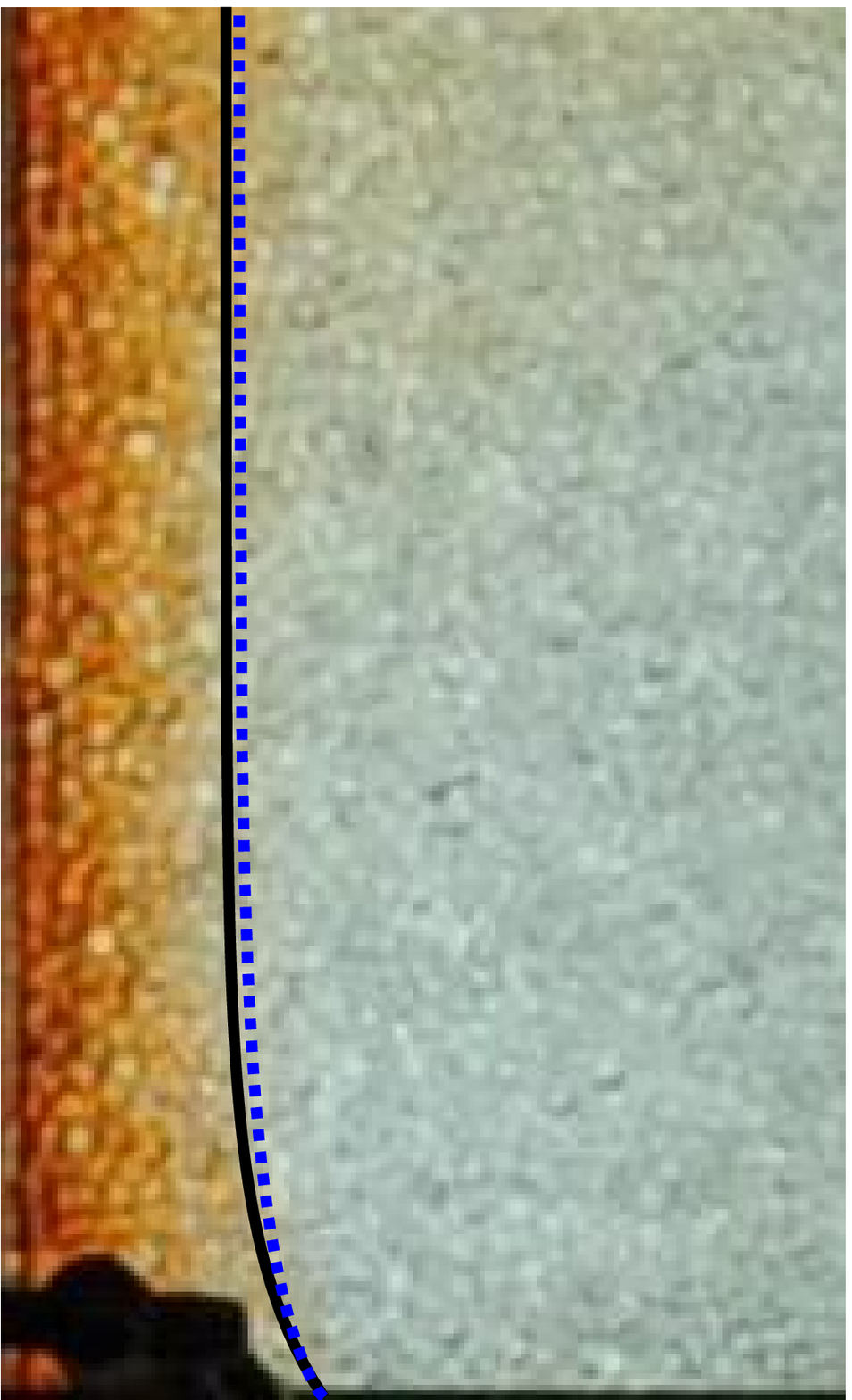}};
\node at (0.6,2) {(a)};
\node at (4.1,2) {(b)};
\draw[line width =2,->] (-1,-1.5) -- (-1,1.5);
\draw[line width =1,->] (-0.55,-1.75) -- (1.8,-1.75);
\draw[line width =1,->] (-0.55,-1.75) -- (-0.55,2.);
\draw[line width =1,->] (2.95,-1.75) -- (5.3,-1.75);
\draw[line width =1,->] (2.95,-1.75) -- (2.95,2.);
\node at (1.9,-1.9) {$x$};
\node at (5.4,-1.9) {$x$};
\node at (-0.8,2) {$z$};
\node at (2.7,2) {$z$};
\node at (-1.2,0) {\bf \rotatebox{90}{Flow direction}};
\draw[line width =1,<->] (-0.6,-1.9) -- (0.4,-1.9);
\node at (-0.1,-2.1) {$a_0=3$ cm};
\draw[line width =1,<->] (2.9,-1.9) -- (3.8,-1.9);
\node at (3.3,-2.1) {$a_0=2.5$ cm};
\node[white] at (0.8,-0.5) {\bf \scriptsize \rotatebox{52}{Clamp}};
\node[white] at (3.25,-1.6) {\bf \scriptsize \rotatebox{-15}{Clamp}};
\end{tikzpicture}
\caption{Experimental photos (taken from the study of \citet{gilmore2022leakage}) of thinning plumes with $W=0.58$ (a) and $W=0.74$ (b), compared with numerical (solid lines) and analytical (dotted lines) solutions for the steady plume shape in the case of a line source. The photos are partially obscured by a clamp (part of the apparatus) which is labelled for clarity.  \label{exppic}}
\end{figure}

The focus of their study was on the leakage of salty water through a gap in an impermeable division midway down the tank.
However, for comparison with the present study we focus on the flow below this division only and we ignore all the flow details above this. 
Therefore, we restrict our attention to the lower $40\times40\times 1\mathrm{cm}$ of their tank. In this way, the leakage rates of salty water into this lower section of the tank (which were calculated in their study), correspond with the injection flow rate $Q$ in our theoretical model. As described by \citet{gilmore2022leakage} both the leakage flux and the near-field plume shape within this lower section of the tank were approximately steady (after an initial transient). Hence, for comparison with our model a constant inflow $Q$ and a steady shape $\alpha(Z)$ can be assumed to good approximation.

Examples of steady plumes from the study of \citet{gilmore2022leakage} are shown in experimental photographs in figure \ref{exppic}a,b.
The plume width at the inlet for each case is $a_0=3,2.5$ cm, respectively. 
To calculate the velocity ratios $W$ for each case, we estimate the far-field plume width downstream $a_\infty$, noting that $W=a_\infty/a_0$. This results in $W=0.58$ and $W=0.74$ for figure \ref{exppic}a and b, respectively.
The steady plume shape predicted by our numerical model is compared to each of these photos with solid lines. The analytical approximation, given by integrating \eqref{plumedif}, is also shown with dotted lines. Overall, good agreement is observed between the numerical model, the analytical approximation, and the experiments. Dispersion causes the plume shape to slightly diffuse downstream of the inlet, which is not captured by the sharp interface in our model.

One advantage of our simple model is that it only depends on a single dimensionless parameter, $W$, which is easily calculated by estimating the plume width at two locations (i.e. $a_0$ and $a_\infty$). A more complicated model which accounts for dispersion, for example, would require further parameter values of the fluid-medium properties, such as the diffusion and dispersion coefficients of the salt/dye.

\subsection{Plume shape: The case of a circular source}
\label{subsec_rad}

In the case of a circular source the boundary conditions \eqref{bcfirst}-\eqref{bclast} are replaced by corresponding conditions in cylindrical radial coordinates (e.g. with $x$ replaced by $r$ and $u$ by $u_r$, the radial velocity).
In this case the radius of the plume (measured from the $z$ axis) is given by $a(z)=(A(z)/\pi)^{1/2}$. As before, we seek a solution of the form \eqref{form1}-\eqref{form3} (with $u$ replaced by $u_r$).
The pressure $\hat{p}$ satisfies a similar system of equations to \eqref{redfirst}-\eqref{redlast} with $x$ replaced by $r$. 
In the case where $W$ is close to unity the solution is calculated by separation of variables, giving
\beq
\hat{p}\approx-2(1-W)\Delta \rho g a_0\sum_{n=1}^\infty \frac{J_0(j_{0,n}r/a_0)}{j_{0,n}^2J_1(j_{0,n})} e^{-j_{0,n}z/a_0},
\eeq
where $J_0$ and $J_1$ are the  $0^\mathrm{th}$ and $1^\mathrm{st}$ order Bessel functions of the first kind, and $j_{0,n}$ is the $n^\mathrm{th}$ zero of $J_0$. 
Likewise, the plume shape is given by the solution to
\beq
a'(z)\approx-2(1-W)\sum_{n=1}^\infty \frac{e^{-j_{0,n}z/a_0}}{j_{0,n}}.\label{plumedifrad}
\eeq
In the case where $W$ is not close to unity, we calculate the solution via the numerical method described earlier. After introducing a scaled dimensionless radial coordinate
\beq
R=r/a(z),
\eeq 
Laplace's equation \eqref{redfirst} becomes
\beq
\left[\alpha^{-2}R^{-1}\partial_{R}(R\partial_R)+\lb\partial_Z-R\alpha'\alpha^{-1} \partial_R\rb^2\right]P=0,\label{lapdim_rad}
\eeq
whilst the remaining boundary conditions stay the same as \eqref{dimfirst}-\eqref{dimlast} except with $X$ switched to $R$.
The system of equations is solved using the same finite difference scheme as in Section \ref{subsec_line} (except with the radius truncated at a small but finite value $R=0.01$ to avoid a singular Laplacian). Likewise, a similar determinant function is defined as
\beq
\Delta(Z):=\alpha^2-4\left.P_R\right|_{R=1}^2,\label{detdefrad}
\eeq
which indicates whether or not a real solution exists.

\begin{figure}
\centering
\begin{tikzpicture}
\node at (0,0) {\includegraphics[height=0.3\textheight]{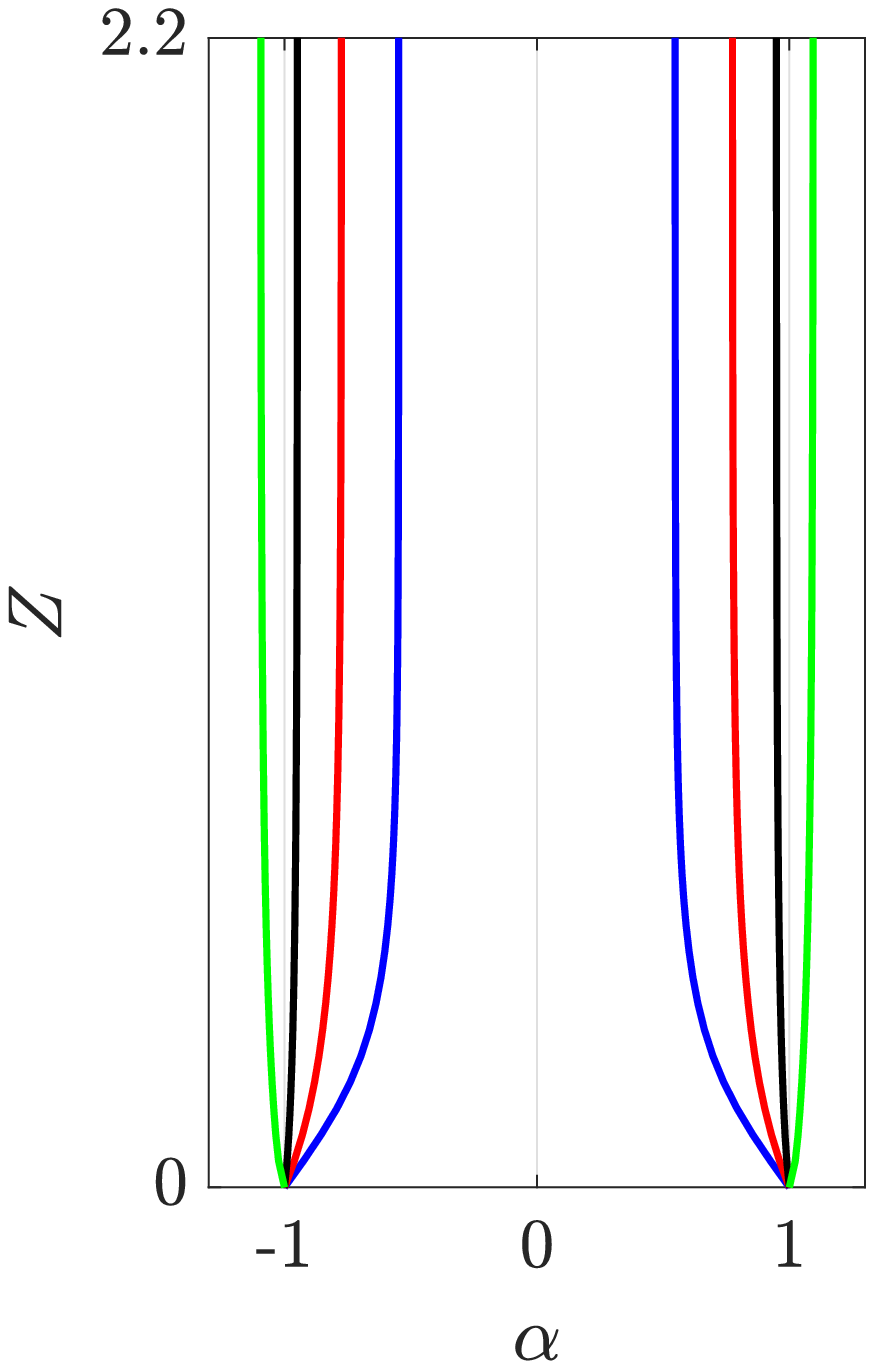}};
\node at (7,0) {\includegraphics[height=0.3\textheight]{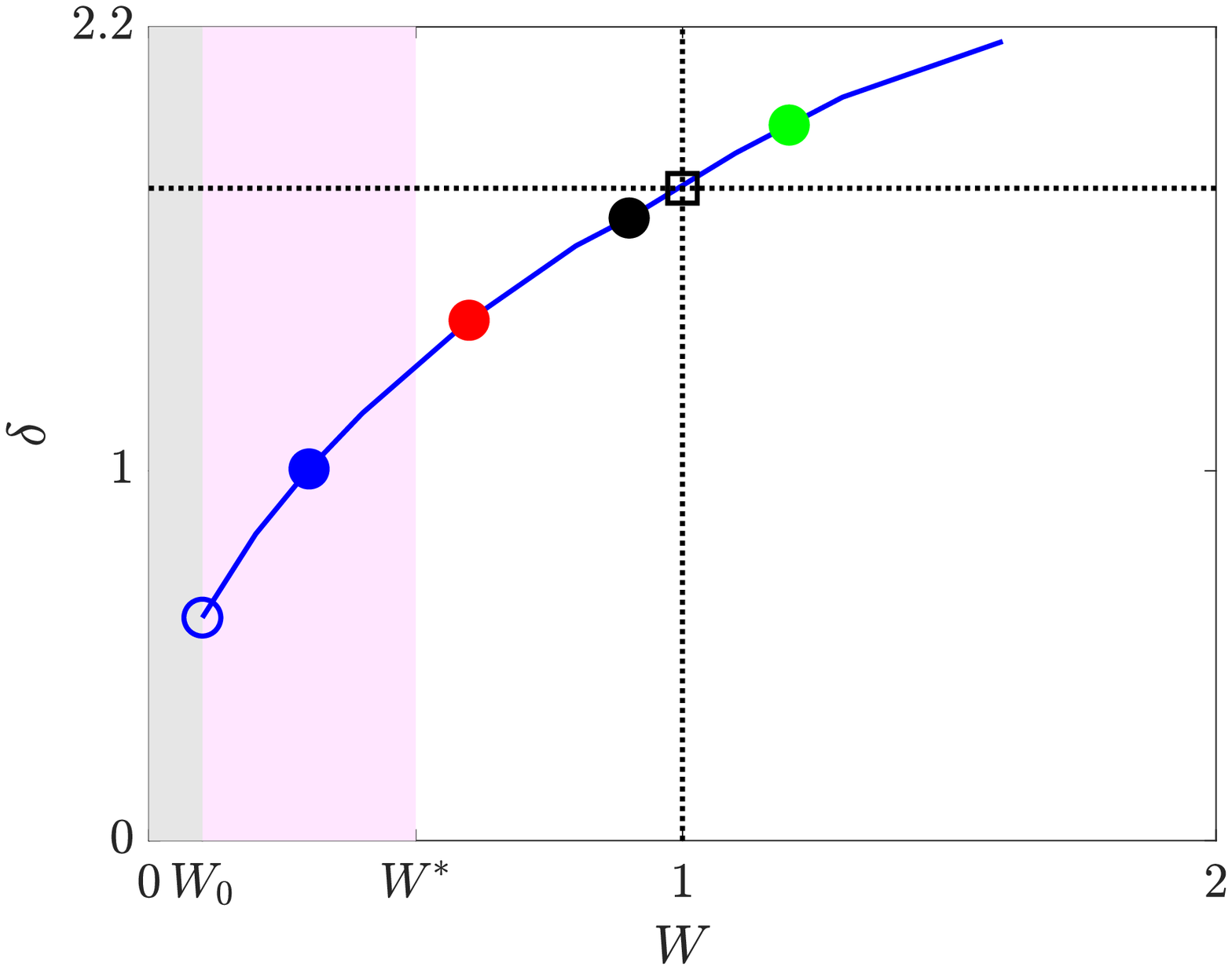}};
\node at (9,-1.5) {\scriptsize Stable};
\node at (4.8,-1.5) {\scriptsize Unstable};
\node at (0.25,3.5) {(a)};
\node at (7,3.5) {(b)};
\node[blue] at (0.6,-1.) {$\delta$};
\draw[blue, line width =1,<->] (0.45,-2.3) -- (0.45,0.2);
\end{tikzpicture}
\caption{Numerical and analytical results for a thinning/thickening axisymmetric plume resulting from a circular source. (a) Plume shape for different velocity ratios $W$ and (b) $99\%$ boundary layer distance $\delta$ as defined in \eqref{deltadef}. Critical values $W_0$ and $W^*$ are related to the sign of the determinant $\Delta$ \eqref{detdefrad} and the stability/existence of a steady solution. Dotted lines indicate the asymptotic limit $W\rightarrow 1$, for which the plume shape is given by the solution to \eqref{plumedifrad}. \label{shape2}}
\end{figure}

In figure \ref{shape2}a we display the steady plume shapes calculated for $W=0.3,0.6,0.9,1.2$. Likewise the $99\%$ boundary layer distance $\delta$ (as defined in \eqref{deltadef}) is plotted in figure \ref{shape2}b. 
Overall the behaviour is similar to the case of a line source, except the plume adjusts over a shorter vertical length scale, resulting in smaller values of $\delta$. 
The approximate solution calculated in the case where $W\approx1$ (i.e. via integration of \eqref{plumedifrad}) results in a boundary layer distance of $\delta\approx 1.76$, as shown with dotted lines in figure \ref{shape2}b. 
Critical values of the velocity ratio (see earlier discussion in Section \ref{subsec_line}) are $W^*=0.5$ and $W_0=0.1$, which are very similar to the case of a line source.


\section{Unsteady plumes and the criteria for stability}
\label{instabsec}

Unlike the previous sections which have assumed a steady state, here we address the possibility of an unsteady flow by investigating the linear stability of the system.
We divide the following analysis into two sections which are distinguished by the velocity ratio $W$. The first section addresses the case when $W_0<W<W^*$ such that the determinant $\Delta$ equals zero at a critical point downstream of the inlet, whilst the second section addresses velocity ratios larger than this, $W>W^*$, for which the determinant is always positive (see discussion at the end of Section \ref{subsec_line}). In the former case we demonstrate the existence of an infinitesimal perturbation to the steady plume shape that grows unbounded over time, and hence we show that such scenarios are inherently unstable. In the latter case we show that such an instability cannot form. 
We ignore the case $W<W_0$ since our numerical method fails to converge to a real-valued solution for such scenarios, leaving us with no base state for the plume. Whilst we focus on the case of the line source in the following sections, the case of the circular source follows approximately the same steps and has similar conclusions.

\subsection{Small velocity ratios $W_0<W<W^*$}\label{subsec_smallW}
 
As described earlier, it is expected that the flow may become unstable for small velocity ratios, such that an unsteady model is required. In the unsteady case, the only equation which requires modification is the kinematic boundary condition \eqref{bcpen}, which becomes
\beq
u=\phi a_t + w a_z:\quad  x=a(z,t),\label{unstead}
\eeq
where $\phi$ is the porosity. Written in terms of the stretched dimensionless coordinates \eqref{stretch_coord}, this becomes
\beq
-\alpha^{-1}P_X=\alpha_T + \left[ 1+ \alpha_Z \alpha^{-1}P_X \right]\alpha_Z:\quad X=1,\label{timedep}
\eeq
where $T=t w_b/a_0\phi $ is the dimensionless time. 
The rest of the governing equations and boundary conditions remain the same as in the steady case \eqref{lapdim}-\eqref{dimmid},\eqref{dimlast}.

We consider a small perturbation applied to the plume shape and pressure of the form
\begin{align}
\alpha&=\bar{\alpha}(Z)+\epsilon \tilde{\alpha}(Z,T),\label{decomp}\\
P&=\bar{P}(X,Z)+\epsilon \tilde{P}(X,Z,T),\label{decomp2}
\end{align}
where $\epsilon\ll1$ is a small parameter and $\bar{\alpha},\bar{P}$ solve the steady problem.  
Inserting \eqref{decomp},\eqref{decomp2} into \eqref{timedep} and linearising, we get 
\beq
\tilde{\alpha}_T+(1+2\bar{\beta}\bar{P}_X)\tilde{\alpha}_Z -\bar{P}_X(\bar{\beta}^2+\bar{\alpha}^{-2})  \tilde{\alpha} +(\bar{\beta}^2\bar{\alpha}+\bar{\alpha}^{-1})\tilde{P}_X=0:\quad X=1,\label{operat}
\eeq
where we have introduced the notation $\bar{\beta}=\bar{\alpha}'/\bar{\alpha}$. 
In addition to \eqref{operat} we require a set of equations and boundary conditions for the perturbed pressure $\tilde{P}$ to complete the system. These are placed in Appendix \ref{app_pres} for convenience. 

\begin{figure}
\centering
\begin{tikzpicture}
\node at (0,0) {\includegraphics[height=0.29\textheight]{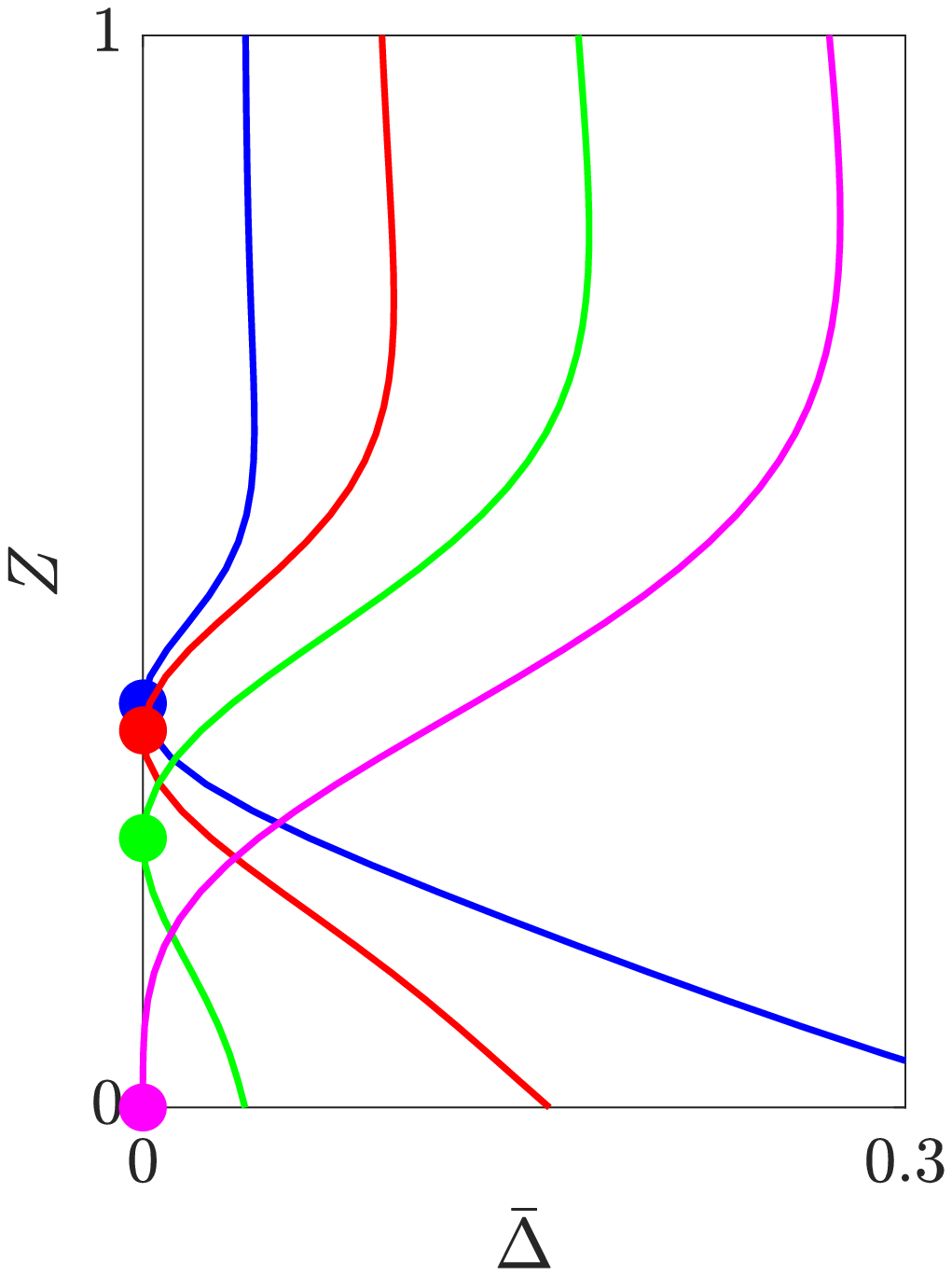}};
\node at (7,0) {\includegraphics[height=0.29\textheight]{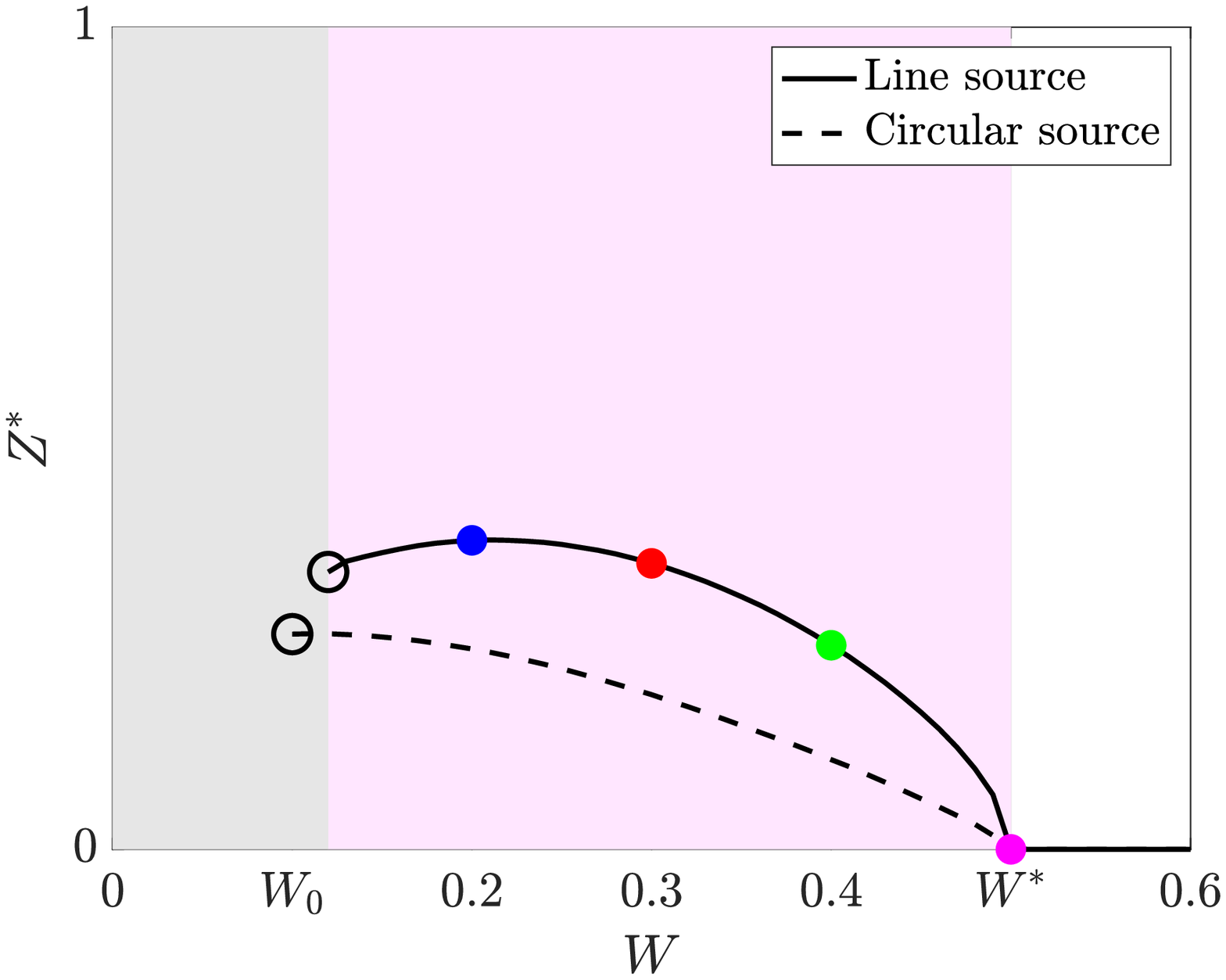}};
\node at (7.,0.5) {\scriptsize Unstable};
\node at (10.15,0.5) {\scriptsize Stable};
\node at (0.25,3.2) {(a)};
\node at (7,3.2) {(b)};
\end{tikzpicture}
\caption{(a) Determinant function $\Delta$ and (b) vertical position of the critical point $Z^*$ where the determinant equals zero, in the case of a line source. The values of $Z^*$ calculated in the case of a circular source are shown in (b) with a dashed line. \label{zstar}}
\end{figure}

The stability of the system is elucidated by considering the determinant function \eqref{detfun} at leading order $\bar{\Delta}=\bar{\alpha}^2-4\bar{P}_X^2$.
As described earlier, for small values of the velocity ratio $W_0<W<W^*$, the determinant function becomes zero at a critical point, $Z=Z^*$, downstream of the inlet. 
To illustrate this we have plotted the determinant function in figure \ref{zstar}a for $W=0.2,0.3,0.4,0.5$, and the corresponding critical points $Z^*$ are plotted alongside in figure \ref{zstar}b.   
Clearly $\bar{\Delta}$ is a non-monotone function that touches zero just once, with critical values in the range $Z^*\in[0,0.37]$ and $Z^*=0$ at $W=W^*$ (also note the corresponding values for the case of a circular source, shown as a dashed line).

By definition, at the critical point ($\bar{\Delta}(Z^*)=0$) the shape function $\bar{\alpha}$ and its derivative satisfy 
\begin{align}
\bar{\alpha}(Z^*)&= 2\bar{P}_X(1,Z^*),\label{alphastar}\\
 \bar{\beta}(Z^*)&=-\bar{\alpha}(Z^*)^{-1}.\label{betastar}
\end{align}
Hence, inserting \eqref{alphastar},\eqref{betastar}, into the linearised kinematic condition \eqref{operat} at the critical point $Z=Z^*$ gives the ordinary differential equation
\beq
\bar{\alpha}^*\dot{\tilde{\alpha}}^*=\tilde{\alpha}^*-2\left.\tilde{P}_X^*\right|_{X=1},\label{zstareq}
\eeq
where dots indicate differentiation with respect to time and star superscripts indicate evaluation at the critical point (e.g. $\tilde{\alpha}^*(T)=\tilde{\alpha}(Z^*,T)$).
The stability of the perturbation at the critical point therefore depends on the right hand side of \eqref{zstareq}, which incidentally is proportional to the perturbed horizontal velocity at the edge of the plume ($\tilde{\alpha}^*-2\tilde{P}_X^*=2\bar{\alpha}^*\tilde{\mathcal{U}}^*|_{X=1}$), as shown in Appendix \ref{app_vel}. 
Hence, \eqref{zstareq} is re-written as
\beq
\dot{\tilde{\alpha}}^*=2\left.\tilde{\mathcal{U}}^*\right|_{X=1}.\label{zstareq2}
\eeq
To assess the stability, we consider how the sign of the perturbed velocity $\tilde{\mathcal{U}}|_{X=1}$ relates to the sign of the perturbation $\tilde{\alpha}$ (i.e. whether the shape is deformed inwards or outwards at the critical point).

For positive perturbations $\tilde{\alpha}>0$, the expanded plume shape needs to be filled with fluid, so we expect a net positive velocity perturbation in the vicinity of the critical point, $\tilde{\mathcal{U}}|_{X=1}>0$. Likewise, for negative perturbations $\tilde{\alpha}<0$ a net negative velocity perturbation is required, $\tilde{\mathcal{U}}|_{X=1}<0$, such that the plume can shrink inwards. 
Hence, $\tilde{\mathcal{U}}|_{X=1}$ is positively correlated with $\tilde{\alpha}$, indicating that \eqref{zstareq2} produces unstable solutions that grow unbounded with time when the perturbation is applied close the critical point. 
Note, however, that when the perturbation (and therefore the flow) extends far away from the critical point, it is not obvious how $\tilde{\alpha}$ and $\tilde{\mathcal{U}}|_{X=1}$ are correlated.

\begin{figure}
\centering
\begin{tikzpicture}[scale=1]
\node at (0,0) {\includegraphics[height=0.3\textheight]{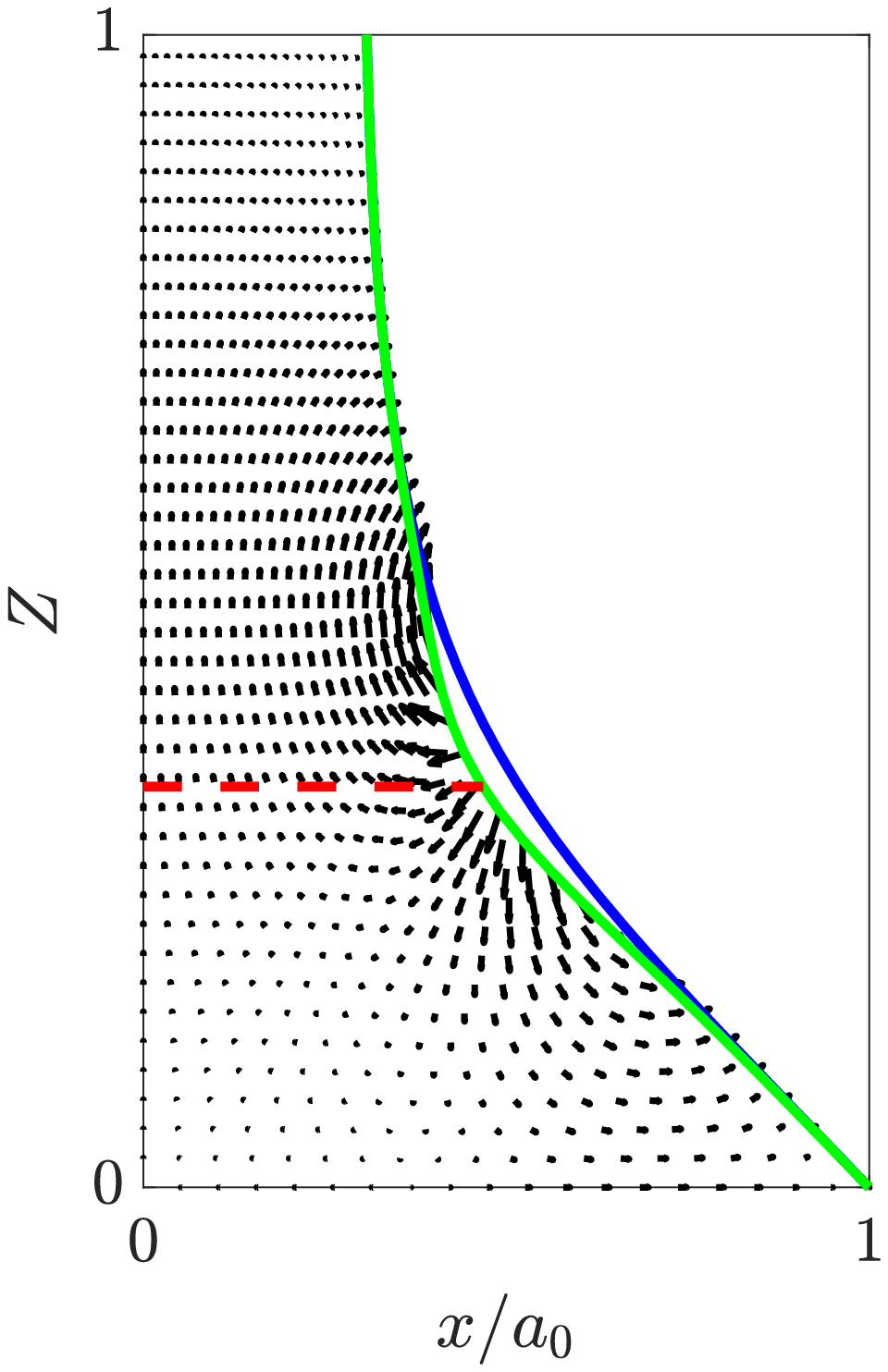}};
\node at (4.5,0) {\includegraphics[height=0.3\textheight]{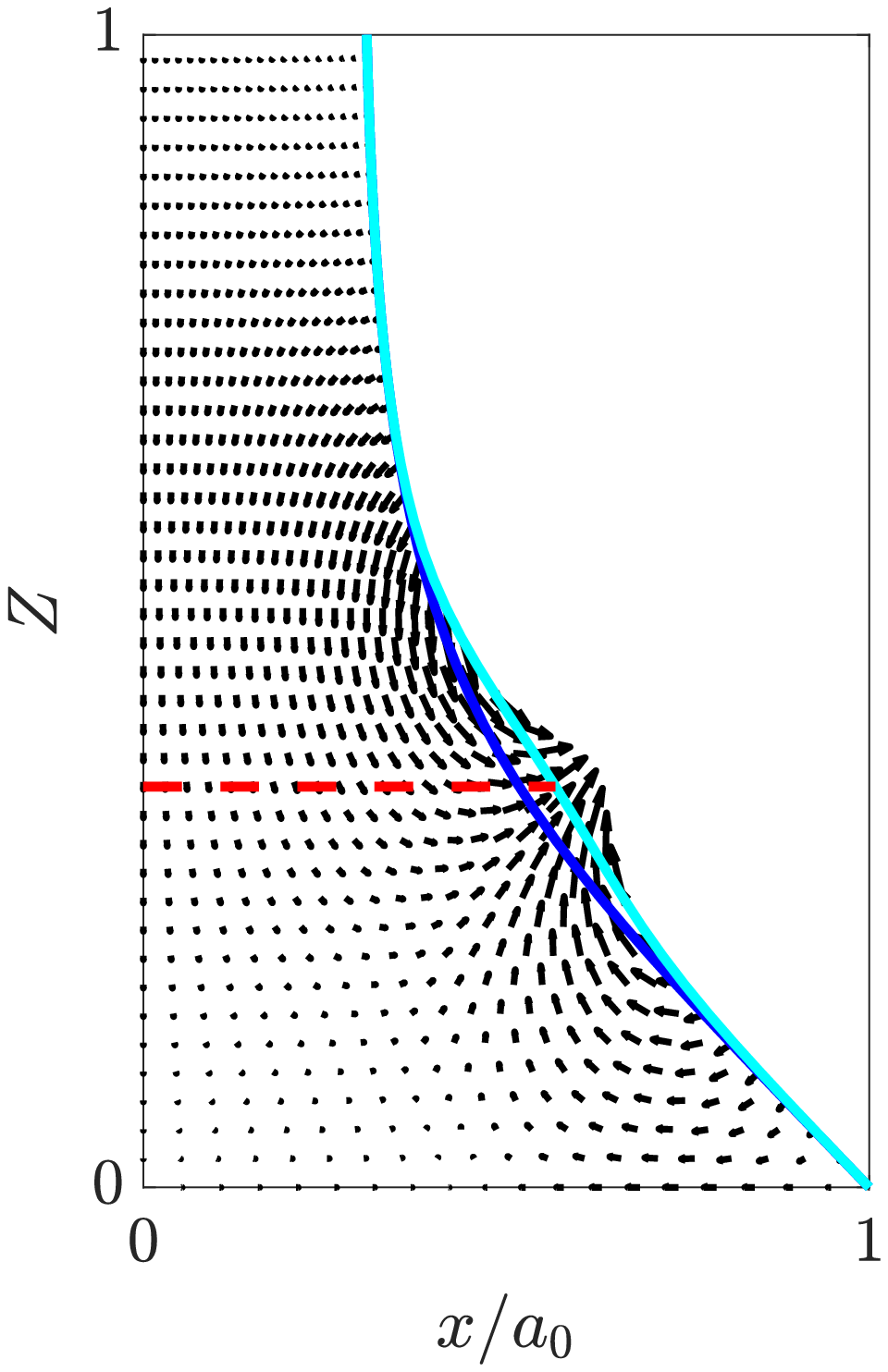}};
\node at (8.8,0) {\includegraphics[height=0.3\textheight]{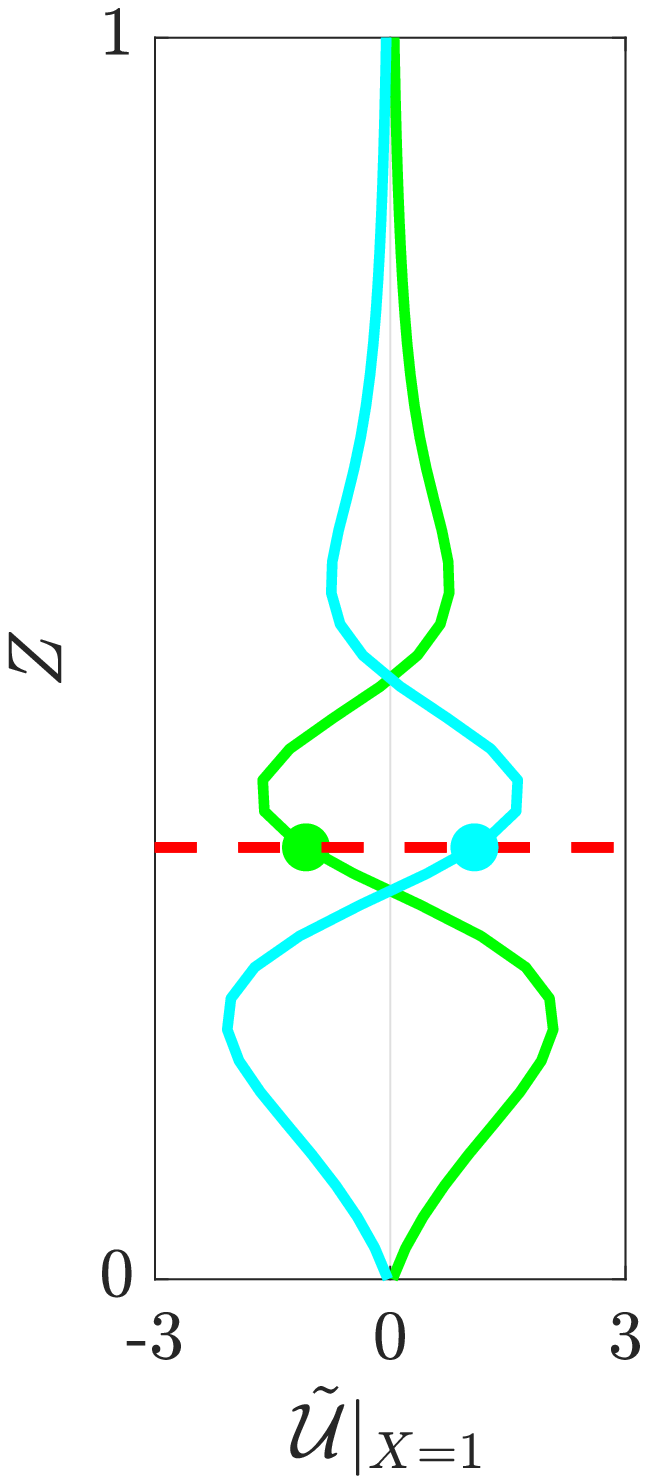}};
\node at (0.3,3.5) {(a)};
\node at (4.8,3.5) {(b)};
\node at (9.0,3.5) {(c)};
\node[green] at (0.7,0.5) {$\boldsymbol{\tilde{\alpha}<0}$};
\node[cyan] at (5.2,0.5) {$\boldsymbol{\tilde{\alpha}>0}$};
\node[red] at (7.8,-0.5) {$\boldsymbol{Z^*}$};
\node[red] at (-1.8,-0.5) {$\boldsymbol{Z^*}$};
\node[red] at (2.8,-0.5) {$\boldsymbol{Z^*}$};
\end{tikzpicture}
\caption{Vector fields for the perturbed velocity, $(\tilde{\mathcal{U}},\tilde{\mathcal{W}})$ (see Appendix \ref{app_vel}), in both the case of a negative perturbation $\tilde{\alpha}<0$ (a) and a positive perturbation $\tilde{\alpha}>0$ (b). The stability is determined by the sign of the perturbed horizontal velocity at the edge of the plume $\tilde{\mathcal{U}}|_{X=1}$ which is plotted in (c) for each case. The steady state plume shape is indicated with a solid blue line in (a),(b). The velocity ratio for this case is $W=0.3$ which has a critical point at $Z^*=0.35$.
\label{schem2}}
\end{figure}

Next we demonstrate the existence of a small localised perturbation that grows unbounded over time. To do so we choose a simple Gaussian function for the initial perturbation, which is of the form
\beq
\tilde{\alpha}(Z^*,0)=\exp\left[ -{(Z-Z^*)^2}/{2\sigma^2} \right],\label{gauss1}
\eeq
where $\sigma$ is the standard deviation. Whilst there are many possible local perturbations which engender instability, we use this one since it is simple and demonstrates the point effectively.

We apply the perturbation \eqref{gauss1} (using $\sigma=0.1$) to the case where $W=0.3$ (for which $Z^*=0.35$) and we plot the results in figure \ref{schem2}. Vector fields for the perturbed velocity, $(\tilde{\mathcal{U}},\tilde{\mathcal{W}})$ (see Appendix \ref{app_vel} for full expressions), are plotted in figure \ref{schem2}a,b for both the case of a positive perturbation $\tilde{\alpha}>0$ and a negative perturbation $\tilde{\alpha}<0$. 
As shown by the velocity vectors, positive/negative perturbations result in a local flow outwards/inwards from the steady plume shape. 
Hence, the perturbed horizontal velocity $\tilde{\mathcal{U}}|_{X=1}$, which is plotted in figure \ref{schem2}c, is positive/negative (in the vicinity of the critical point) for positive/negative perturbations, indicating that the solution is unstable.

Whilst this is just a specific case, it nevertheless demonstrates that an infinitesimal  solution can be constructed that becomes unstable.
Although we do not include the results here, we have also developed a time-dependent implicit numerical scheme that solves the perturbed equations for the pressure and plume shape at first order. These time-dependent numerical simulations confirm that localised perturbations of the form \eqref{gauss1} become unstable when applied near the critical point.
A more detailed eigenvalue analysis could explore the fastest growing perturbation (eigenfunction) and corresponding eigenvalue exponent. However, this lies outside the scope of the current study.

\subsection{Moderate velocity ratios $W>W^* \wedge W\approx 1$}\label{subsec_medW}

Now that we have addressed the case of small velocity ratios, for which the determinant becomes zero at a critical point, we next address the case of moderate velocity ratios for which $\Delta$ is always positive. We restrict our attention to velocity ratios which are larger than $W^*$ but which are still order $\mathcal{O}(1)$ (i.e. ignoring $W\gg 1$). Hence, taking $W\approx1$, the pressure and the plume shape are well approximated by the expressions \eqref{gilmoreapprox},\eqref{plumedif} (which are both of order $\mathcal{O}(1-W)$). Meanwhile, the time-dependent kinematic condition \eqref{unstead} approximates to
\beq
-({k}/{\mu})\hat{p}_x\approx \phi a_t + w_b a_z:\quad  x\approx a_0.\label{unstead2}
\eeq
As before, we consider a small perturbation applied to the plume shape and pressure, but now we avoid converting to dimensionless coordinates for simplicity. Hence, we consider a perturbation of the form
\begin{align}
a&=\bar{a}(z)+\epsilon \tilde{a}(z,t),\label{decompmod}\\
\hat{p}&=\bar{p}(x,z)+\epsilon \tilde{p}(x,z,t).\label{decompmod2}
\end{align}
Attention is required when performing this decomposition, since there are now two small parameters in the problem, namely $\epsilon>0$ and $\varepsilon=1-W>0$ (we consider $W<1$ without loss of generality).
Hence, in the following analysis it is assumed that the perturbation to the shape is relatively much smaller than the perturbation to uniform flow, such that an asymptotic hierarchy $0<\epsilon\ll\varepsilon\ll 1$ is maintained.

After inserting \eqref{decompmod},\eqref{decompmod2} into \eqref{redfirst}-\eqref{redpen},\eqref{redlast},\eqref{unstead2},  and expanding in powers of $\epsilon$, it is clear that the leading order expressions for the pressure and shape, $\bar{p}$, $\bar{a}$, are precisely \eqref{gilmoreapprox},\eqref{plumedif}. Meanwhile, the unsteady terms satisfy
\begin{align}
-({k}/{\mu})\tilde{p}_x=\phi\tilde{a}_t+w_b\tilde{a}_z:\quad  x\approx a_0,\label{finaleq}
\end{align}
and the pressure perturbation $\tilde{p}$ satisfies homogeneous versions of the governing equations and boundary conditions \eqref{redfirst}-\eqref{redpen},\eqref{redlast} (i.e. with zero on the right hand side of all the equations). Hence, the pressure perturbation is trivial $\tilde{p}=0$ and consequently \eqref{finaleq} also becomes homogeneous. In this way, perturbations to the plume shape $\tilde{a}$ are simply advected downstream with dimensional velocity $w_b/\phi$, and consequently the system is stable.  

We have therefore shown that plume shapes are stable in the regime $W\approx1$, and unstable in the regime $W_0<W<W^*$. It is not known whether the plume shapes for $W>W^*$ but $W\not\approx1$ are stable or unstable (e.g. one could argue that $W=0.6\not\approx1$), and for this a full eigenvalue analysis is required. 
Such an analysis could predict the largest value of $W$ that onsets instability (i.e. producing a positive real-valued eigenvalue).
However, for the purposes of this study, and since our experimental data suggests a stable shape for velocity ratios as small as $W=0.58$ (see figure \ref{exppic}a), this analysis covers the relevant and interesting range of cases.

To go beyond the current analysis and resolve the nonlinear instability at small velocity ratios, a time-dependent numerical simulation is required either in two or three dimensions (e.g. see \citet{hewitt2013stability,hewitt2017stability}). 
In particular, it is not clear exactly how the flow evolves over time, whether it forms fingers, filaments or disconnected regions. 
Moreover, in the presence of dispersion, it is likely that thin disconnected regions may fuse together.  
Hence, due to the complications associated with such a numerical simulation, this lies outside the scope of the current study.

\section{Concluding remarks}

We have studied the shape and stability of buoyancy-driven plumes near their injection point within a porous medium.  
The key controlling parameter is the ratio between the inlet velocity and the far-field (buoyancy) velocity. 
Whether this ratio is larger or smaller than one determines whether the plume is expanding or contracting downstream. 
For small values of this ratio, the plume shape becomes unstable at a critical point downstream, which we have demonstrated using a linear stability analysis. 
On the other hand, when the velocity ratio is close to one, we have shown that the plume shape is stable. 

Future work could include the effects of mixing between the injected fluid and the ambient fluid, similarly to \citet{sahu2015filling,lyu2016experimental}. In particular, as the two fluids mix together, the difference in density between them becomes smaller. Hence, buoyancy decreases and the width of the plume increases as the flow moves downstream. As described earlier in Section \ref{sec_plumeintro}, the experiments of  \citet{sahu2015filling} indicated that mixing takes place over vertical length scales $\sim 10\times$ larger than the plume width, whereas we have calculated values of $\delta$ (i.e. shape adjustments) within length scales close to $\sim 1\times$ the plume width (e.g. see figures \ref{shape} and \ref{shape2}). Hence, we have confirmed that dispersion occurs at larger length scales than the adjustments to the plume shape studied here in the near-field.

It would also be interesting to investigate the possible effects of a time-varying injection $Q(t)$ \citep{huppert1982propagation}. In particular, it may be possible to stabilise/destabilise the near-field plume shape by dynamically controlling the flow rate appropriately. In this case, the timescale for changes in flow rate scales like $t\sim Q/\dot{Q}$, whereas the timescale for buoyancy is $t\sim a_0\phi/w_b$. Hence, we require sufficiently large values of $a_0\phi\dot{Q}/w_bQ$ to be able to regulate the flow in this way.

It is worth noting the relevance of this study to the case of CO$_2$ sequestration, in which buoyant CO$_2$ (originating from emissions due to power plants and industrial processes) is injected into subsurface geological aquifers for storage. 
In this case, low permeability sedimentary layers may cause the CO$_2$ to spread out as it rises \citep{bickle2007modelling,cowton2016inverse,cowton2018benchmarking}.
It is important to be able to quantify the behaviour of the buoyant CO$_2$ if and when it penetrates through these low permeability layers and migrates further upwards. Hence, our study provides a tool for modelling the flow behaviour near the breakthrough locations (as well as near injection points, in general), which is useful for quantifying leakage rates \citep{gilmore2022leakage} and the consequent CO$_2$ migration speeds \citep{neufeld2009effect}. Moreover, the stability of the plume determines how it mixes with the surrounding brine (i.e. by breaking apart into filaments, fingers or disconnected regions), thereby influencing how it becomes trapped due to dissolution and residual trapping \citep{nordbotten2011geological,krevor2015capillary}. 
However, in such cases there are other more complicated physical phenomena to account for, such as the dissolution of the CO$_2$ \citep{macminn2012spreading}, multiphase effects \citep{golding2011two} and flow rearrangement due to geological  heterogeneities \citep{benham2021two,benham2021upscaling}.

\acknowledgements{
The author would like to thank Kirean A Gilmore for supplying the experimental photographs and further information regarding the experiments.}\\

Declaration of Interests. The author reports no conflict of interest.\\

\appendix

\section{First order pressure perturbation}
\label{app_pres}

In this section we briefly describe the system of equations required to calculate the first order pressure perturbation, which is used to analyse stability in Section \ref{subsec_smallW}. 
After inserting the decomposition \eqref{decomp}-\eqref{decomp2} into the governing equation for the pressure \eqref{lapdim} and linearising in terms of the small paramter $\epsilon\ll1$, the corrected equation at first order is
\beq
\begin{split}
A_{XX}\tilde{P}_{XX} +A_{XZ}\tilde{P}_{XZ}+ A_{ZZ} \tilde{P}_{ZZ}  + A_X\tilde{P}_X   =B_{ZZ}\tilde{\alpha}_{ZZ} +B_Z\tilde{\alpha}_Z + B\tilde{\alpha},
 \end{split}
\eeq
where the coefficients (using subscript notation for clarity) are given by
\begin{align}
A_{XX}&=\bar{\alpha}^{-1}(1 + \bar{\alpha}^2  \bar{\beta}^2  X^2),\\
A_{XZ}&=- 2  \bar{\alpha}  \bar{\beta} X,\\
A_{ZZ}&= \bar{\alpha},\\
A_X&=\bar{\alpha}X (\bar{\beta}^2  -   \bar{\alpha} \bar{\beta}' ),\\
B_{ZZ}&=\bar{P}_{X} X,\\
B_Z&=-2  X (2  \bar{\beta} \bar{P}_X - \bar{P}_{XZ} +  \bar{\beta} \bar{P}_{XX} X),\\
B&=-\bar{\alpha}^{-2} \left[\bar{\alpha}^2 (-3  \bar{\beta}^2 \bar{P}_X +  \bar{\beta}' \bar{P}_X + 2 \bar{\beta} \bar{P}_{XZ}) X - 
      2 \bar{P}_{XX} (1 + \bar{\alpha}^2  \bar{\beta}^2 X^2)\right].
\end{align}
Likewise, the boundary conditions \eqref{dimfirst}-\eqref{dimmid},\eqref{dimlast}, linearised and keeping only first order terms, become
\begin{align}
\tilde{P}_X&=0:&X=0,\\
\tilde{P}_Z-X\bar{\beta} \tilde{P}_X&=\bar{P}_X X\bar{\alpha}^{-1}( \tilde{\alpha}_Z-\bar{\beta}\tilde{\alpha}):&Z=0,\\
\tilde{P}_Z-X\bar{\beta} \tilde{P}_X&\rightarrow \bar{P}_X X\bar{\alpha}^{-1}( \tilde{\alpha}_Z-\bar{\beta}\tilde{\alpha}):&Z\rightarrow \infty,\\
 \tilde{P}&=0:&X=1.
\end{align}
The new system of equations are solved together (using the same finite difference scheme described earlier) with the linearised kinematic boundary condition \eqref{operat} to acquire the first order pressure and shape $\tilde{P},\tilde{\alpha}$.

\section{First order velocity perturbation}
\label{app_vel}

In this section we briefly derive expressions for the perturbed velocity field which are used to create the vector fields in figure \ref{schem2}. 
Following the same notation used in Section \ref{subsec_line}, the dimensionless flow velocities are given by 
\begin{align}
\hat{u}/w_b&=- \alpha^{-1}{P}_X,\label{vel_pert_app1}\\
\hat{w}/w_b&=-{P}_Z+X\alpha'\alpha^{-1}{P}_X.\label{vel_pert_app2}
\end{align}
Next, we insert \eqref{decomp} and \eqref{decomp2} into \eqref{vel_pert_app1}-\eqref{vel_pert_app2}, and expand in powers of $\epsilon$, keeping only leading order and first order terms. We denote the leading order velocities as $\bar{\mathcal{U}},\bar{\mathcal{W}}$, and the first order velocities as  $\tilde{\mathcal{U}},\tilde{\mathcal{W}}$. These are given by
\begin{align}
\bar{\mathcal{U}}&=- \bar{\alpha}^{-1}\bar{P}_X,\\
\bar{\mathcal{W}}&=-\bar{P}_Z+X\bar{\beta}\bar{P}_X,\\
\tilde{\mathcal{U}}&=\bar{\alpha}^{-2}\left[\tilde{\alpha}\bar{P}_X-\bar{\alpha}\tilde{P}_X\right],\label{velpertass1}\\
\tilde{\mathcal{W}}&=-\tilde{P}_Z + \left[(-\bar{\beta}\tilde{\alpha} + \tilde{\alpha}_Z) \bar{P}_X\bar{\alpha}^{-1} + \bar{\beta} \tilde{P}_X\right] X.\label{velpertass2}
\end{align}
The final two expressions \eqref{velpertass1}-\eqref{velpertass2} are precisely the terms used to plot the perturbed velocity vector field in figure \ref{schem2}.

It should be noted that at the critical point, $X=1$, $Z=Z^*$, the above equations simplify to
\begin{align}
\bar{\mathcal{U}}&=- 1/2,\\
\bar{\mathcal{W}}&=-1/2,\\
\tilde{\mathcal{U}}&=(2\bar{\alpha})^{-1}\left[\tilde{\alpha}-2\tilde{P}_X\right],\label{velpertass12}\\
\tilde{\mathcal{W}}&= (\bar{\alpha}^{-1}\tilde{\alpha} + \tilde{\alpha}_Z)/2 -\bar{\alpha}^{-1} \tilde{P}_X.\label{velpertass22}
\end{align}
Hence, we see that the horizontal velocity perturbation \eqref{velpertass12} is proportional to the right hand side of \eqref{zstareq}.

\bibliographystyle{jfm}
\bibliography{bibfile.bib}

\end{document}